\RequirePackage[T1]{fontenc}
\documentclass[12pt]{article}

\usepackage[height=8.85in,width=6.45in]{geometry}

\usepackage[utf8]{inputenc}
\usepackage{amsmath}
\usepackage{amssymb}
\usepackage{mathtools}
\numberwithin{equation}{section}
\usepackage{slashed}
\usepackage{braket}
\usepackage[svgnames]{xcolor}
\usepackage[colorlinks,citecolor=DarkGreen,linkcolor=FireBrick]{hyperref}
\usepackage{cite}
\usepackage{graphicx}
\usepackage{tikz}
\usepackage{tikz-cd}
\usepackage{times}
\usepackage{courier}
\usepackage{bm}
\usepackage{subfig}

\usepackage{xcolor}
\usepackage{mdframed}

\renewenvironment{figure}[1][]{
  \begin{originalfigure}[#1]
    \begin{mdframed}[linecolor=black!0,backgroundcolor=black!0]
}{
    \end{mdframed}
  \end{originalfigure}
}

\usepackage{draft}
\usepackage{youngtab}
\def\Nequals#1{$\mathcal{N}{=}#1$}
\def\SO{\mathrm{SO}}
\def\SU{\mathrm{SU}}
\def\SL{\mathrm{SL}}
\def\U{\mathrm{U}}
\def\USp{\mathrm{USp}}
\def\diag{\mathrm{diag}}
\begin{document}

\begin{titlepage}

\begin{flushright}
IPMU-18-0066
\\
PUPT-2561
\end{flushright}

\vskip 3cm

\begin{center}

{\Large \bfseries Comments on the twisted punctures of $A_\text{even}$ class S theory}

\vskip 1cm
Yuji Tachikawa$^1$, Yifan Wang$^2$ and Gabi Zafrir$^1$
\vskip 1cm

\begin{tabular}{ll}
$^1$ & Kavli Institute for the Physics and Mathematics of the Universe, \\
& University of Tokyo,  Kashiwa, Chiba 277-8583, Japan\\
$^2$ & Joseph Henry Laboratories, Princeton University, \\
&Princeton, NJ 08544, USA \\
\end{tabular}

\vskip 1cm

\end{center}

\noindent
We point out that the $\USp$ symmetry associated to a full twisted puncture of a class S theory of type $A_\text{even}$ has the global anomaly associated to $\pi_4(\USp)=\mZ_2$.
We discuss manifestations of this fact in the context of the superconformal field theory  $R_{2,2N}$ introduced by Chacaltana, Distler and Trimm.
For example, we find that this theory can be thought of as a natural ultraviolet completion of an infrared-free $\SO(2N+1)$ gauge theory with $2N$ flavors, whose $\USp(4N)$ symmetry clearly has the global anomaly.

\end{titlepage}

\setcounter{tocdepth}{1}
\tableofcontents


\section{Introduction and summary}
The class S theory, introduced by Gaiotto and Gaiotto-Moore-Neitzke in \cite{Gaiotto:2009we,Gaiotto:2009hg},  is a large class of 4d \Nequals2 superconformal field theory obtained by
compactifying 6d \Nequals{(2,0)} theory on a Riemann surface with punctures.
This construction not only allowed a geometric understanding of various S-dualities,
but also provided a huge variety of new 4d \Nequals2 theories.
This variety comes from various sources: we have the choice of the initial 6d \Nequals{(2,0)} theory,
which comes with the ADE classification;
then one can introduce punctures, which come roughly in two varieties, called the regular ones and the irregular ones;
then one can further introduce twists by outer automorphisms. 

The class S theories with regular punctures have been systematically explored by Chacaltana, Distler and their collaborators \cite{Chacaltana:2010ks,Chacaltana:2011ze,Chacaltana:2012zy,Chacaltana:2012ch,Chacaltana:2013oka,Chacaltana:2014jba,Chacaltana:2014ica,Chacaltana:2014nya,Chacaltana:2015bna,Chacaltana:2016shw,Chacaltana:2017boe,Distler:2017xba,Chcaltana:2018zag},
for almost all possible types of 6d \Nequals{(2,0)} theories with outer automorphism twists.
The remaining two cases are the twisted punctures of type $A_\text{even}$ theories,
and the case where $\bZ_2$-twisted and $\bZ_3$-twisted punctures of type $D_4$ are combined.
The aim of this paper is to make a small comment on the former case, namely the twisted punctures of type $A_\text{even}$ theories.
More specifically, we point out that \emph{the flavor symmetry $\USp(2N)$ of the full twisted puncture of $A_{2N}$ theory has the global anomaly associated to $\pi_4(\USp(2N))=\mZ_2$.}

This point can be seen most succinctly as follows.
Consider splitting a Riemann surface on which the 6d \Nequals{(2,0)} theory of type $A_{2N}$ is compactified along a long tube around which we have $\bZ_2$ outer-automorphism twist.
This results in two twisted full punctures coupled by the 5d theory obtained by compactifying the $A_{2N}$ theory on $S^1$ with $\bZ_2$ outer-automorphism twist.
This is the maximally supersymmetric 5d $\USp(2N)$ theory with the discrete theta angle $\theta=\pi$ \cite{Hori:1998iv,Tachikawa:2011ch}.
The 4d class S theory with a twisted full puncture, in this viewpoint, 
lives on a boundary of this 5d $\USp(2N)$ theory.

First, this determines that the current algebra central charge of $\USp(2N)$ symmetry of the twisted full puncture is $k=2N+2$ in the normalization where the half-hypermultiplet in the fundamental has $k=1$.\footnote{We emphasize that the flavor central charge here of the twisted full puncture is valid for class S constructions {\it without} irregular punctures. Adding an irregular puncture will modify the $\U(1)_{ R}$ symmetry of the 4d theory and thus change the flavor central charge which is related to the $\U(1)_{R}$ anomaly \cite{Shapere:2008zf}.}
More importantly for us, this means that the $\USp(2N)$ symmetry of the twisted full puncture has the global anomaly.
This is because of the following.
Note that the discrete theta angle is controlled by $\pi_4(\USp(2N))=\bZ_2$,
which also controls the global anomaly on the 4d $\USp(2N)$ symmetry, as originally discussed by Witten \cite{Witten:1982fp}.
Therefore, there is an anomaly inflow from the bulk to the boundary,
and the twisted full puncture needs to carry the global anomaly.
This is analogous to the fact that if the bulk 5d theory has $\SU(n)$ gauge symmetry with the level $k$ Chern-Simons term,
the boundary 4d theory which is coupled to this bulk 5d theory should have 't Hooft anomaly corresponding to $k$ chiral  Weyl fermions in the fundamental representation of $\SU(n)$.

Note that the twisted full puncture of class S theory of type $D_{N+1}$ also has the symmetry $\USp(2N)$, with the same current algebra central charge $k=2N+2$.
The essential difference here is that this puncture does not have the global anomaly.

In this paper, we discuss various manifestations of this fact,
mainly using the 4d \Nequals2 superconformal field theory $R_{2,2N}$ introduced in \cite[Sec.~7.2]{Chacaltana:2012ch} for $N=1$ and in \cite{Chacaltana:2014nya} for general $N$.
This theory is obtained by compactifying the 6d \Nequals2 theory of type $A_{2N}$ on a sphere with one simple puncture and two twisted full punctures.
In \cite{Chacaltana:2014nya}, it was shown that the symmetry $\USp(2N)^2$ apparent in this construction enhances to $\USp(4N)$,
but Witten's anomaly could not be directly determined, since the diagonal combination $\USp(2N)_\text{diag}\subset \USp(2N)^2\subset \USp(4N)$ was gauged.
Our main observation is that by turning on the mass term associated to the simple puncture,
the theory becomes an $\SO(2N+1)$ theory with $2N$ hypermultiplets in the vector representation, which is infrared free;
the $\USp(4N)$ symmetry of this theory clearly has the global anomaly.

The rest of the paper consists of three sections, which can be read independently, and are written using different techniques.
Namely, the section~\ref{sec:classS} uses the class S construction,
the section~\ref{sec:5d} uses the dimensional reduction from 5d,
and the section~\ref{sec:orient} uses a very traditional orientifold construction.

In Sec.~\ref{sec:classS}, after briefly reviewing the original construction of $R_{2,2N}$, we provide a different construction of the same $R_{2,2N}$ theory as a sphere of $A_{4N}$ theory with a full twisted puncture and an irregular puncture, generalizing the construction of \cite{Wang:2015mra}.
This allows us to perform a consistency check of the global anomaly.

In Sec.~\ref{sec:5d},
we point out that the mass deformation for the $\U(1)$ flavor symmetry associated to the simple puncture gives rise to the $\SO(2N+1)$ theory coupled to $2N$ flavors. 
This will be done by constructing these theories by a twisted compactification of  5d \Nequals1 theory, generalizing the work of \cite{Zafrir:2016wkk}.

In Sec.~\ref{sec:orient},
we revisit this mass deformation from the point of view of the old-fashioned type IIA construction.
We will see that the standard Seiberg-Witten curve of the $\SO(2N+1)$ theory with $2N$ flavors, in the standard MQCD form, 
is literally equal to the Seiberg-Witten curve of the $R_{2,2N}$ theory,
such that the dynamical scale $\Lambda$ of the $\SO(2N+1)$ theory is simply the mass parameter of the $R_{2,2N}$ theory.

\section{Class S constructions}
\label{sec:classS}
\subsection{Review of $R_{2,2N}$} 
\begin{figure}[htb]
	\centering
	\includegraphics[scale=.4]{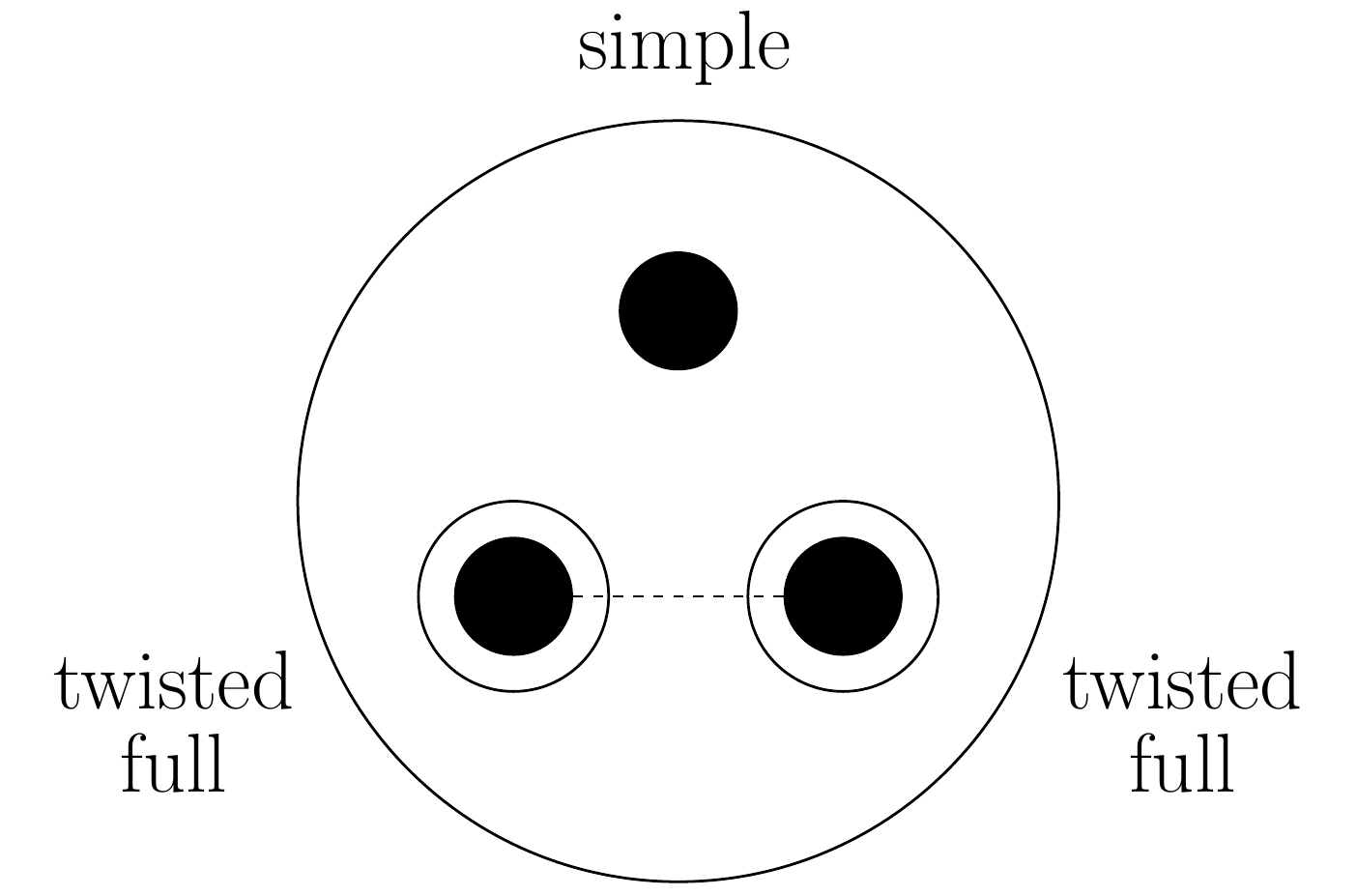}
	\caption{The three-punctured sphere for the theory $R_{2,2N}$.}
	\label{foo}
\end{figure}

Let us first recall how the $R_{2,2N}$ non-Lagrangian SCFT was constructed in \cite{Chacaltana:2014nya}. We start with a class S theory of $A_{2N}$ type with the UV curve given by a torus with minimal untwisted puncture decorated by an $\mZ_2$ twist line along a handle of the torus. 
The theory has two S-dual frames with gauge theory descriptions:
\ie
\Yvcentermath1 \yng(2)-\SU(2N+1)-\Yvcentermath1 \yng(1,1)
\quad
 \mathrel{\mathop{\rightleftarrows}^{\mathrm{S-dual}}_{\mathrm{}}}
 \quad
 \USp(2N)- R_{2,2N}
 \label{sdf}
\fe  
and the $R_{2,2N}$ SCFT emerges from the weak coupling limit of $\USp(2N)$ gauge coupling. Motivated by the decoupling picture, one can engineer this SCFT directly using 3 regular $A_{2N}$ punctures on a sphere: one minimal untwisted puncture, and two maximal twisted punctures, see Fig.~\ref{foo}.

The decoupling picture above determines the Coulomb branch spectrum of $R_{2,2N}$ SCFT to be
\ie
 \Delta\in\{3,5,7,\dots, 2N+1\}
 \fe
 and the conformal central charges are
 \ie
 a={14N^2+19 N+1\over 24},\quad c={8N^2+10N+1\over 12}.
 \fe
 The theory has enhanced $\U(1)\times \USp(4N)_{2N+2}$ flavor symmetry where the $\USp(4N)$ factor has the flavor central charge $k_{\USp(4N)}=2N+2$. Consequently, the 2d chiral algebra of the SCFT in the sense of \cite{Beem:2013sza} naturally contains the affine Kac-Moody algebra of type $C_{2N}$ with level $k_{2d}=-(N+1)$ and a weight one current realizing the 4d $\U(1)$ symmetry. Furthermore $k_{\USp(4N)}$  saturates a flavor central charge unitary bound of \cite{Beem:2013sza},
which means that the stress tensor in the 2d chiral algebra is given by the Sugawara construction   \cite{Chacaltana:2014nya}.\footnote{We emphasize here that the full chiral algebra of the $R_{2,2N}$ SCFT contains additional Virasoro primaries than those generated by the current algebra, as evident from the Hall-Littlewood index of the theory \cite{Chacaltana:2014nya}.}
 
The flavor symmetry  $\USp(4N)$ is enhanced from its maximal subgroup $\USp(2N)\times \USp(2N)$ which is manifest from the twisted maximal punctures. In \eqref{sdf}, diagonal subgroup $\USp(2N)$ is conformally gauged. 

As argued in the introduction,  the twisted $A_{2N}$ maximal punctures which carry $\USp(2N)$ flavor symmetries should have a $\mZ_2$-valued  global anomaly. This would be consistent if the $\USp(4N)$ symmetry of $R_{2,2N}$ also carries the global anomaly; then the $\USp(2N)$ diagonal gauging is non-anomalous. Thus if we have an alternative construction $R_{2,2N}$ that makes the enhanced flavor symmetry manifest as a twisted puncture, this would offer a nontrivial consistency check. We show this is indeed the case in the next section.
Our alternative construction involves irregular punctures.

\subsection{Class S theories of type $A_{2N}$ with a twisted irregular puncture}
\begin{figure}[htb]
	\centering
	\includegraphics[scale=.4]{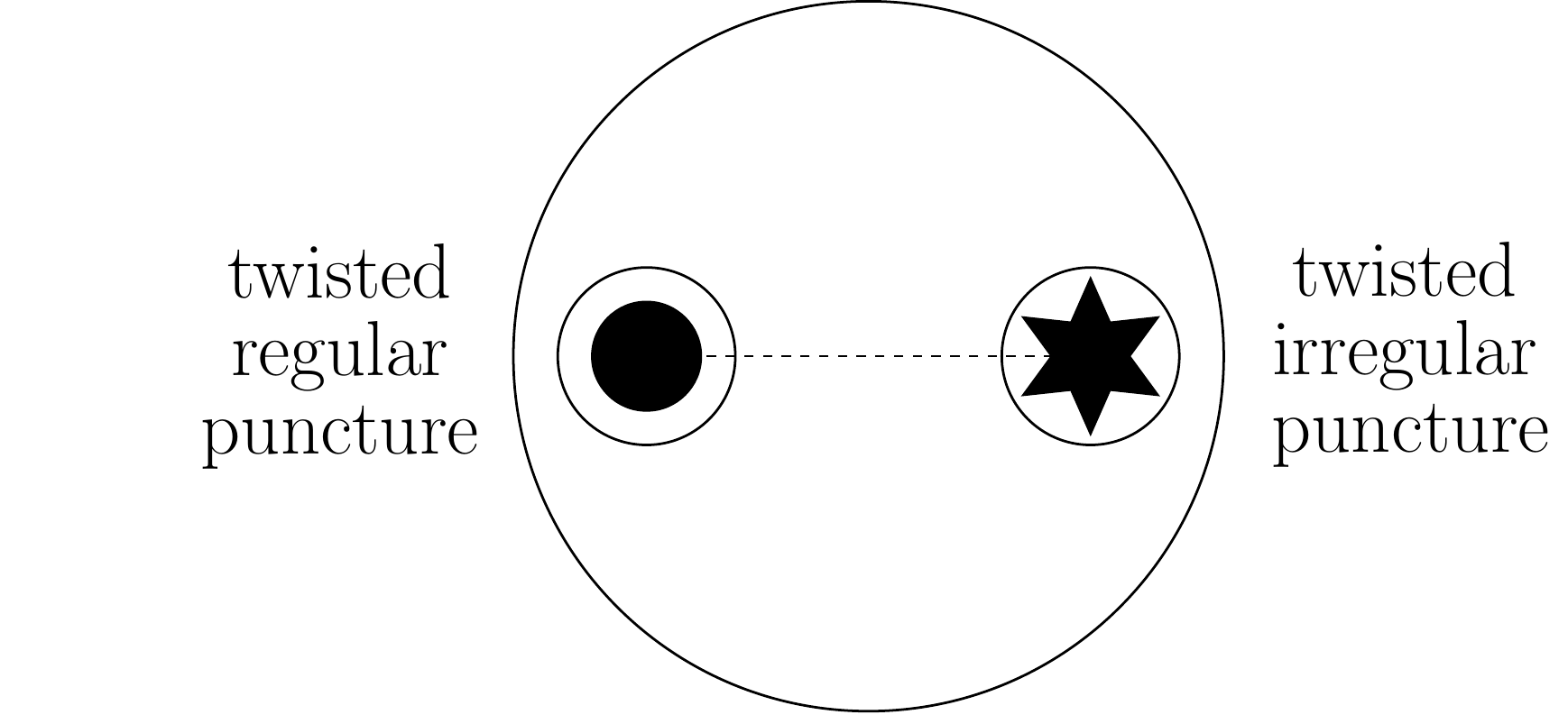}
	\caption{The sphere with one twisted regular puncture and one twisted irregular puncture that engineers a Class S SCFT of type $A_{2N}$ with $\mZ_2$ twist.}
	\label{fig:irreg}
\end{figure}

Consider general 4d $\cN=2$ SCFTs in class S of type $A_{2N}$  from one regular maximal twisted puncture and one irregular twisted  puncture, see Fig.~\ref{fig:irreg}.

The twisted punctures correspond to codimension-2 defects in the (2,0) theory of $A_{2N}$ type and are specified by a local singularity of the Higgs field $\Phi$ at $z=0$ on the UV curve. For the Higgs field to be well-defined, we require 
\ie
\Phi(e^{2\pi i} z)=g[o(\Phi(z))]g^{-1}
\label{gi}
\fe
as one circles the singularity. Here $o$ denotes the $\mZ_2$ outer-automorphism associated to the $A_{2N}$ Dynkin diagram and $g$ generates an inner automorphism of $A_{2N}$.

Among the twisted punctures, regular punctures are associated with simple poles for $\Phi$ and are classified in \cite{Chacaltana:2012zy}, while the irregular punctures involve higher order poles and a detailed classification will appear in \cite{wangxie}. For our purpose here, we list two distinguished classes of twisted irregular punctures of $A_{2N}$ type and the relevant physical data from \cite{wangxie} below:

\paragraph{Class I} has the Higgs field of the form
 \ie
 \Phi={ T \over z^{2+{\kappa\over 4N+2}}}+\cdots
 \label{hp1}
 \fe
 for $\kappa$ odd, where  
\ie
  T=\diag(1,\omega^2,\omega^4,\ldots,\omega^{4N}),\qquad \omega^{4N+2}=1.
\fe
 We have the conjectural flavor central charge formula
  \ie
k_{\USp(2N)}=2N+2-{1\over 2}{4N+2\over 4N+2+\kappa}.
 \label{fcA2N1}
 \fe

 \paragraph{Class II} has the Higgs field of the form
 \ie
 \Phi={ T \over z^{2+{\kappa\over 2N}}}+\cdots
  \label{hp2}
 \fe
 where  
\ie
 T=\diag(0,1,\omega,\ldots,\omega^{2N-1}),\qquad \omega^{2N}=1.
\fe
 We have the conjectural flavor central charge formula
  \ie
k_{\USp(2N)}=2N+2-{1\over 2}{2N\over 2N+\kappa}.
 \label{fcA2N2}
 \fe

The conformal central charges for the corresponding 4d SCFTs are determined by
\ie
2a-c={1\over 4}\sum (2[u_i]-1),~~c={1\over 24}{k_G  \dim(G) \over -k_G/2+h^{\vee}}-{f\over 12}.
\label{conjcc}
\fe
Here $f$ is the number of mass parameters coming from the irregular singularity which equals \textit{one} for the Class II case and \textit{zero} for the Class I case. $k_G$ is the flavor symmetry central charge in \eqref{fcA2N1} and \eqref{fcA2N2}.
 
The conjectured central charge formulae are for the two parameter family of theories labelled by $(N,\kappa)$ either in Class I or II. 
We provide extensive nontrivial pieces of evidence for the above conjectures in  \cite{wangxie}.

\subsection{$R_{2,2N}$ from  twisted irregular punctures}
\label{fromtp}
Given the general construction of 4d $\cN=2$ SCFTs from twisted irregular punctures in the last section, we now construct $R_{2,2N}$ from 6d $(2,0)$ $A_{4N}$ type theory on a sphere with two twisted punctures: one regular maximal puncture, and one irregular puncture of class II with $\kappa= 1-4N$ in \eqref{hp2}. 
The irregular twisted puncture is described by the following local singularity for the Higgs field $\Phi$,
 \ie
 \Phi={  T \over z^{2+{1-4N\over 4N}}}+\cdots
 \fe
 where  
\ie
  T=\diag(0,1,\omega,\ldots,\omega^{4N-1}),\qquad \omega^{4N}=1.
\fe
This is chosen such that under $z\to z e^{2\pi i}$, the Higgs field transform by an automorphism of order $4N$ that is a product of the $\mZ_2$ outer-automorphism $o$ associated to the $A_{4N}$ Dynkin diagram and an inner automorphism of $A_{4N}$ as in \eqref{gi}.

 The singular Seiberg-Witten curve is
 \ie
x^{4N+1}+xz^{1-4N} =0
\fe
 and the Seiberg-Witten 1-form is $\lambda= xdz$. The scaling dimensions of the coordinates are
 \ie
 {}[x]=1-4N,\quad [z]=4N.
 \fe
 Under the $\mZ_2$ twist, the differentials $\phi_\ell$ transform as
 \ie
 \phi_\ell(z) \to (-1)^\ell \phi_\ell(z).
 \fe
 Therefore $\phi_\ell$ for $\ell$ even (odd) involves integral (half-integral) powers of $z$.  We can immediately read off the spectrum of Coulomb branch operators to be
 \ie
 \Delta=\{3,5,7,\dots, 2N+1\}
 \fe
which all come from the odd-degree differentials $\phi_\ell$ with $\ell=2N+3,2N+5,\dots, 4N+1$. 

The $\USp(4N)$ flavor symmetry comes from the regular twisted puncture and its flavor central charge is determined by \eqref{fcA2N2} to be $2N+2$. The differential $\phi_{2N+1}$ contributes the additional mass parameter responsible for the  $\U(1)$ factor in the flavor symmetry.

 It is also easy to check that the central charges computed from \eqref{conjcc}  and \eqref{fcA2N2} are consistent with the results in the previous section. 
Moreover we can directly see  that the Seiberg-Witten curves from the two descriptions agree. For example let us look at the $N=1$ case which is a rank-1 theory with a Coulomb branch operator $u$ of dimension 3. In this case $[x]=-3$ and $[z]=4$. The full Seiberg-Witten curve in the $A_4$ description is
\ie
x^5+x^3{\epsilon_2\over z^2}
+
x^2 {m\over z^{5\over 2}}
+x 
{\epsilon_4+z\over z^4}  
+ 
{u\over z^{9\over 2}}=0.
\fe
On the other hand, the full Seiberg-Witten curve in the $A_2$ description is 
\ie
x^3+ x{\xi_2 +m^2z+\xi'_2 z^2\over z^2(z-1)^2}
-{i u\over z^{5\over 2}(z-1)}=0
\fe
with $[x]=1$ and $[z]=0$. In both cases the Seiberg-Witten differential is taken to be $\lambda=xdz$. Here $\epsilon_2,\epsilon_4$ label Casimirs of the $\USp(4)$ flavor symmetries and $\xi_2,\xi'_2$ label Casimirs of the $\SU(2)\times \SU(2)$ subgroup. The additional $\U(1)$ mass is labelled by $m$. For simplicity, let us look at the curves with the mass parameters turned off:
\ie
A_4:x^5+{x 
	\over z^3} +
{u\over z^{9\over 2}}=0,
\qquad 
A_2:x^3-
{i u\over z^{5\over 2}(z-1)}=0.
\fe
Starting from the $A_4$ curve, we perform a coordinate redefinition $z\to -z^3 x^4,~x\to {1\over x^3 z^2}$ which only changes the Seiberg-Witten differential by an exact 1-form, and then the $A_4$ curve becomes identical to the $A_2$ curve, after throwing out irrelevant overall factors.

\subsection{Half-hypermultiplet from irregular twisted puncture}

Since we expect Witten's global anomaly to be a local property of the twisted full puncture, which is eventually a boundary condition for the 5d super Yang-Mills theory, 
it should not depend on the types of  other punctures that are involved in a given Class S setup. In other words, if we can tune the choice of the irregular puncture in Fig.~\ref{fig:irreg} such that the 4d theory has a free/weakly-coupled description that clearly exhibits the anomaly, it strongly indicates that the twisted full puncture carries the global anomaly. Below we see this is indeed the case for a  free half-hyper multiplet in the fundamental representation of $\USp(2N)$.  

The half-hyper can be constructed using $A_{2N}$ twisted punctures: a Class I twisted irregular puncture with $\kappa=1-(4N+2)$ in \eqref{hp1}
\ie
\Phi={ T\over z^{1+{1 \over   4N+2}}}+\cdots 
\fe
and one twisted full  regular puncture with $\USp(2N)$ flavor symmetry. The spectral curve  is
\ie
x^{2N+1}+\sum_{i=1}^{N}z^{2i} \epsilon_{2i} +z^{-{1\over 2}-2N}=0.
\fe
Here the scaling dimensions are determined by $[x]=-1-4N$ and $[z]=4N+2$. Thus $\epsilon_{2i}$ with dimension $2i$ is a degree $2i$ Casimir for the $\USp(2N)$ flavor symmetry.  The above is also consistent with \eqref{fcA2N1} which gives
\ie
k_{\USp(2N)}=1
\fe
as expected for a half-hyper.

\section{5d constructions}
\label{sec:5d}

In the previous section we relied on class S methods to argue that the $R_{2,2N}$ SCFT has a global  anomaly for the $\USp(4N)$ group. Here we shall show further evidence for this by using a different realization of the $R_{2,2N}$ SCFT. The realization we employ is the compactification of 5d SCFTs with a global symmetry twist. In this manner, 4d $\mathcal{N}=2$ SCFTs can be engineered by the compactification of 5d SCFTs. This method can be used to engineer many 4d $\mathcal{N}=2$ SCFTs including non-Lagrangian theories appearing in class S constructions\cite{Benini:2009aa}.

Many 5d SCFTs possess discrete global symmetries. It is then possible to consider a compactification with a twist under said discrete symmetry. In other words the compactification is done such that upon traversing the circle one comes back to the theory acted by the discrete symmetry element. To try to understand the results of such compactifications it is useful to consider 5d SCFTs with a string theory construction that exhibits the global symmetry.

A convenient way to realize 5d SCFTs in string theory is using brane webs\cite{AH,AHK}. Discrete symmetries of the SCFTs are then manifested by discrete symmetries of the brane system. A particular interesting case is when the symmetry is manifested on the web as a combination of spacetime reflections and an $\SL(2,\mZ)$ transformation. The cases where the discrete symmetry is $\mZ_2$ and $\mZ_3$ were analyzed in \cite{Zafrir:2016wkk} where it was argued that such a construction can realize the $R_{2,2N}$ SCFT. We next review some aspects of this construction that will be important for us.

\begin{figure}
\center
\includegraphics[width=0.85\textwidth]{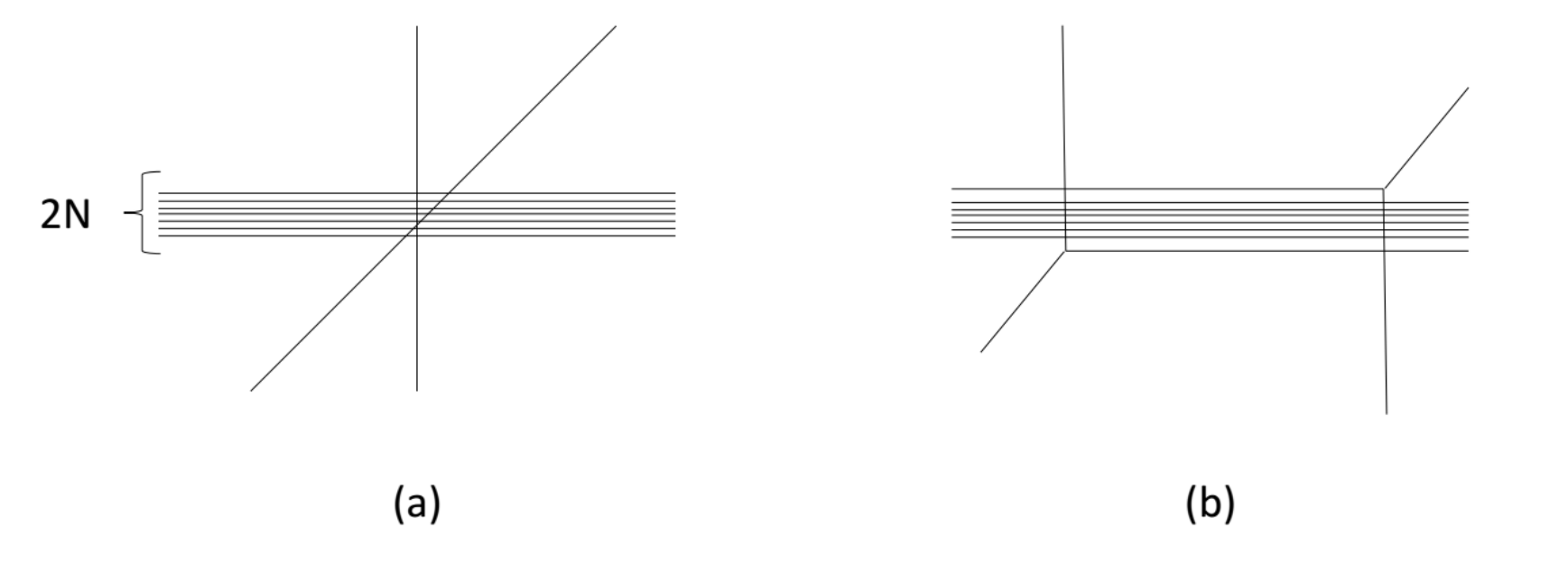} 
\caption{(a) The brane web for a 5d SCFT with global symmetry $\SU(4N)\times \U(1)^2$. (b) The same web after moving some of the $7$-branes. This corresponds to a mass deformation of the 5d SCFT. As can be seen from the web, this mass deformation sends the 5d SCFT to an $\SU(2N+1)$ gauge theory with $4N$ flavors in the fundamental representation.}
\label{WebTw}
\end{figure}

Consider the bane web shown in figure \ref{WebTw} (a). This describes a 5d SCFT which lives at the intersection of all the $5$-branes. This SCFT can be defined in field theory as the UV fixed point of an $\SU(2N+1)$ gauge theory with $4N$ flavors in the fundamental representation. This can be seen from figure \ref{WebTw} (b) which shows the web after a mass deformation corresponding to the $\SU(2N+1)$ coupling constant.  

This 5d SCFT has an $\SU(4N)\times \U(1)^2$ global symmetry. Additionally, it also has a $\mZ_2$ discrete symmetry, which in the brane web is given by a $\pi$ rotation in the plane of the web combined with  $-I$ in the $\SL(2,\mZ)$  transformation. In the low-energy $\SU(2N+1)$ gauge theory it is manifested as charge conjugation which is a symmetry of the gauge theory. 

We can now consider compactifying the 5d SCFT with a twist under this discrete symmetry. It was argued in \cite{Zafrir:2016wkk} that this should give the 4d $R_{2,2N}$ SCFT. There are several reasons for this identification. One is that they have the same symmetries. Here the $\SU(4N)$ part is projected to $\USp(4N)$ by the twist while one $\U(1)$ remains and another is projected out. This is readily seen in the 5d gauge theory description by considering how charge conjugation acts on these symmetries. Another reason is that one can argue from the brane web that the resulting theory needs to participate in the same duality (\ref{sdf}).

We can now use the brane construction to study various properties of the $R_{2,2N}$ SCFT. The particular properties, that are of interest to us here, are mass deformations. We have already encountered one such mass deformation, the one leading to the $\SU(2N+1)$ gauge theory. This deformation is invariant under the $\mZ_2$ discrete symmetry as can be seen from the low-energy gauge theory. We thus expect it to remain also in the $R_{2,2N}$ SCFT where it should correspond to a deformation in the $\U(1)$ global symmetry.

We can infer the result of this deformation from the 5d construction, where it should just be the twisted compactification of the 5d gauge theory. As the twist acts on it by charge conjugation, we expect the $\SU(2N+1)$ to be projected to $\SO(2N+1)$ while the $4N$ flavors should be split to two groups of $2N$ each mapping to the other. The end result is a 4d $\SO(2N+1)$ gauge theory with $2N$ hypermultiplets in the vector representation. This theory indeed has a $\USp(4N)$ global symmetry as required. Furthermore it is easy to see that it has a global Witten's anomaly. By anomaly matching arguments this implies that the starting theory, the $R_{2,2N}$ SCFT, must also have the same anomaly.   

\begin{figure}
\center
\includegraphics[width=0.85\textwidth]{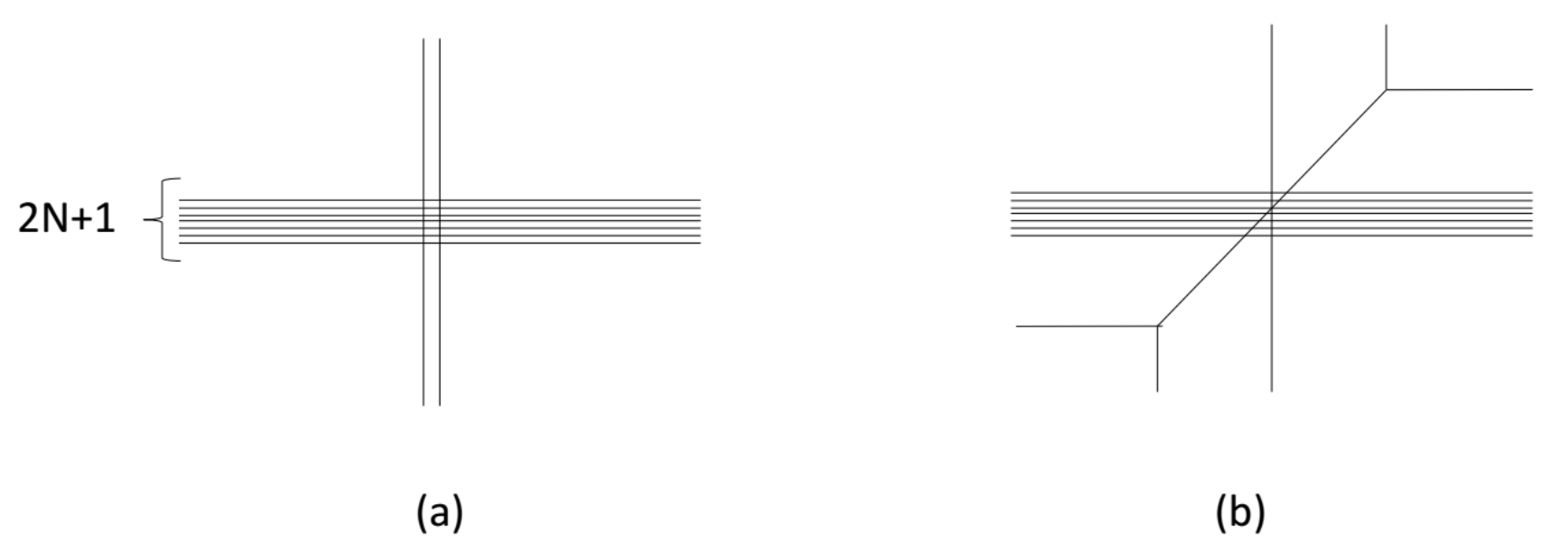} 
\caption{(a) The brane web for a 5d SCFT with global symmetry $\SU(4N+2)\times \SU(2)^2$. (b) The same web after moving some of the $7$-branes. This corresponds to a mass deformation of the 5d SCFT. As can be seen from the web, this mass deformation sends this 5d SCFT to the 5d SCFT in figure \ref{WebTw} (a).}
\label{WebTw1}
\end{figure}

We can also consider mass deformations leading to the $R_{2,2N}$ SCFT. For instances consider the 5d SCFT shown in figure \ref{WebTw1} (a). This is a 5d SCFT that has the same discrete symmetry and so we can also consider its compactification to 4d with a twist. The result of such a compactification was considered in \cite{Zafrir:2016wkk} where it was argued that it leads to a known 4d SCFT with $\SU(2)\times \USp(4N+2)$ global symmetry. An interesting property of this 4d SCFT is that it is dual to an $\SO(2N+3)$ gauge theory with $2N+1$ hypermultiplets in the vector representation upon gauging its $\SU(2)$ global symmetry. This duality in particular implies that its $\USp(4N+2)$ global symmetry has a global anomaly.
This theory can also  be engineered by a Class S of $A_{4N+2}$ type involving one twisted irregular puncture and one twisted regular full puncture that realizes the $\USp(4N+2)$ flavor symmetry with the global anomaly \cite{wangxie}.

 As shown in figure \ref{WebTw1} (b) we can flow from this SCFT to the $R_{2,2N}$ SCFT via a mass deformation. Then anomaly matching arguments again suggest that the $\USp(4N)$ global symmetry of the $R_{2,2N}$ SCFT has a global anomaly.

\section{IIA constructions}
\label{sec:orient}
\begin{figure}
\centering
\begin{tabular}{cc}
\includegraphics[scale=.5]{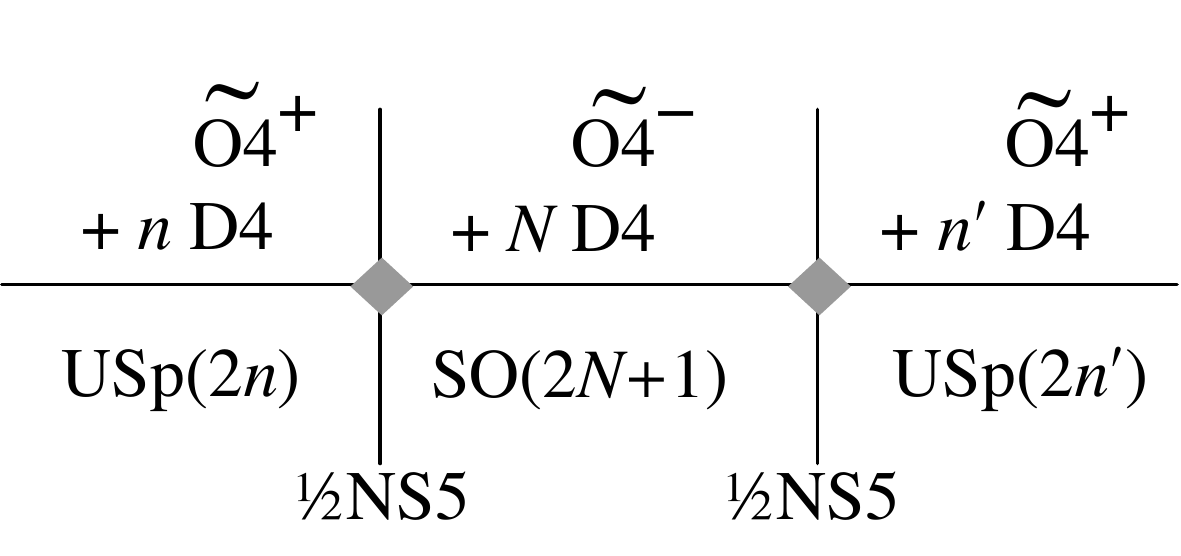} & \includegraphics[scale=.5]{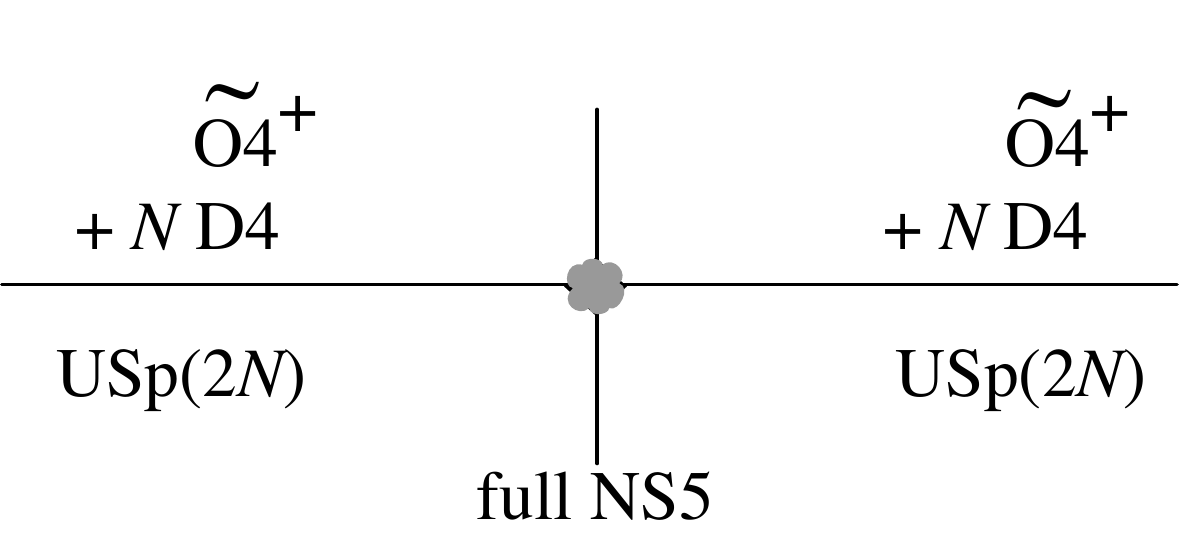}\\
(a) & (b)
\end{tabular}
\caption{(a) The IIA brane setup to engineer  $\SO(2N+1)$ gauge theories with $n+n'$ flavors. 
This requires the use of two half-NS5-branes. 
(b) When we collapse the two half-NS5-branes, we obtain the IIA reduction for the $R_{2,2N}$ theory.}
\label{o}
\end{figure}

In this section we provide another way to see that the $R_{2,2N}$ SCFT can be mass deformed to be the $\SO(2N+1)$ gauge theory with $2N$ flavors, using a traditional brane construction \cite{Witten:1997sc} using orientifolds \cite{Hori:1998iv,Gimon:1998be}. 
Let us first recall the brane construction of 4d $\SO(2N+1)$ gauge theories with hypermultiplets in the vector representation using O4, D4 and NS5 branes, see Fig.~\ref{o} (a).

As shown in the figure, this requires the use of $\widetilde{\mathrm{O4}}^-$-plane to realize the $\SO(2N+1)$ gauge group.
Then, across a half-NS5-brane, the type of the orientifold plane switches to $\widetilde{\mathrm{O4}}^+$-plane,
and we have a half-hypermultiplet in the bifundamental of $\USp$ and $\SO(2N+1)$ at the intersection of the D4-branes and the half-NS5-brane.
Note that this bifundamental half-hypermultiplet has the global anomaly for $\USp$ symmetry.
From the anomaly inflow argument, this implies that the 5d $\USp$ gauge symmetry on the $\widetilde{\mathrm{O4}}^+$-plane has to have the discrete theta angle $\theta=\pi$. 
This is a much simpler argument for this theta angle than the one given in \cite{Tachikawa:2011ch}.

Instead, consider reducing the three-punctured sphere defining the $R_{2,2N}$ theory given in Fig.~\ref{foo} from M-theory to type IIA.
A twisted full puncture corresponds to a semi-infinite segment of $N$ D4-branes on the $\widetilde{\mathrm{O4}}^+$-plane,
and the untwisted simple puncture corresponds to a full NS5 brane, see Fig.~\ref{o} (b).
Clearly, the two setups shown in Fig.~\ref{o} (a) and (b) are related by the motion of the two half-NS5-branes,
when $n=n'=N$.
Since the distance between the two half-NS5-branes specify the squared inverse gauge coupling of the $\SO(2N+1)$ gauge theory at the string scale,
this clearly means that the $R_{2,2N}$ theory is at the infinite coupling limit of the $\SO(2N+1)$ gauge theory with $2N$ flavors. 

What remains is to show that how the gauge coupling looks like from the point of view of the $R_{2,2N}$ theory.
Here we encounter a mild surprise: the Seiberg-Witten curve of the $\SO(2N+1)$ theory with $2N$ flavors is \emph{the same} with the Seiberg-Witten curve of the $R_{2,2N}$ theory, without making any approximation.

To see this, recall the Seiberg-Witten curve of $\SO(2N+1)$ theory with $n+n'$ flavors, as determined by lifting the brane setup shown in Fig.~\ref{o} (a) to M-theory: \begin{equation}
x\left[
\Lambda^{2N-1-2n}\frac1{z^{1/2}}\prod_{i=1}^{n} (x^2-m_i^2) +\Lambda^{2N-1-2n'} z^{1/2} \prod_{i=1}^{n'}(x^2-\tilde m_{i'}^2)
\right]
= x^{2N} + u_2 x^{2N-2} + \cdots + u_{2N}.
\end{equation} with the Seiberg-Witten differential $\lambda=xdz/z$.
This becomes, for $n=n'=N$,  \begin{equation}
x\left[
\frac1{\Lambda z^{1/2}}\prod_{i=1}^{N} (x^2-m_i^2) + \frac{z^{1/2}}{\Lambda} \prod_{i=1}^{N}(x^2-\tilde m_{i'}^2)
\right]
= x^{2N} + u_2 x^{2N-2} + \cdots + u_{2N}.\label{AA}
\end{equation}

We now compare this to the Seiberg-Witten curve of the $R_{2,2N}$ theory. 
We put two full twisted punctures at $z=0$ and $z=\infty$, and the untwisted simple puncture at $z=1$.
The  Seiberg-Witten curve is given by \begin{equation}
x^{2N+1}+ \phi_1(z) x^{2N}+\phi_2(z) x^{2N-1} + \cdots + \phi_{2N+1}(z)=0,
\end{equation} with the Seiberg-Witten differential $\lambda=xdz/z$.
Note that we allow $\phi_1(z)$ to be nonzero to simplify the description of the mass-deformed untwisted simple puncture,
where the conditions just become that for all $k$, $\phi_k(z)$ have at most a simple pole at $z=1$.
Combining with the conditions from the twisted full punctures, we see that the Seiberg-Witten curve of the $R_{2,2N}$ theory to be \begin{equation}
x^{2N+1} + \frac{mz^{1/2}}{z-1} x^{2N}
+ \frac{\mu_2 z + \mu_2'}{z-1} x^{2N-1}
+ \frac{\hat u_3z^{1/2}}{z-1} x^{2N-2}
+ \cdots  
+ \frac{\mu_{2N} z + \mu_{2N}'}{z-1} x
+\frac{\hat u_{2N+1}z^{1/2}}{z-1}=0,\label{BB}
\end{equation}
where $m$ is the mass parameter for the untwisted simple puncture,
$\mu_2,\ldots,\mu_{2N}$ and $\mu'_2,\cdots,\mu'_{2N}$ are the mass parameters for the two twisted full punctures,
and $\hat u_3,\cdots, \hat u_{2N+1}$ are the Coulomb branch parameters of the $R_{2,2N}$ theory.

We see that the curve \eqref{AA} for the $\SO(2N+1)$ theory with $2N$ flavors
and the curve \eqref{BB} for the $R_{2,2N}$ theory are \emph{the same} up to a minor relabeling of the parameters 
as one can see by dividing \eqref{AA} by $(z^{1/2}+z^{-1/2})/\Lambda$ to make the coefficient of $x^{2N+1}$ to be one.
In particular, we see that the scale $\Lambda$ of the Landau pole of the infrared gauge theory can simply be equated 
to the mass deformation parameter $m$ of the $R_{2,2N}$ theory.
It might be interesting to look for similar phenomena in other 4d \Nequals2 gauge theories with infrared-free matter content,
such that its Seiberg-Witten curve is the same as that of some other SCFT.

\section*{Acknowledgments}
YT is partially supported  by JSPS KAKENHI Grant-in-Aid (Wakate-A), No.17H04837 
and JSPS KAKENHI Grant-in-Aid (Kiban-S), No.16H06335,
and also by WPI Initiative, MEXT, Japan at IPMU, the University of Tokyo.
YW is supported in part by the US NSF under Grant No. PHY-1620059 and by the Simons Foundation Grant No. 488653.
GZ is supported in part by World Premier International Research Center Initiative (WPI), MEXT, Japan.

\bibliographystyle{ytphys}
\baselineskip=.95\baselineskip
\bibliography{ref}

\providecommand{\href}[2]{#2}\begingroup\raggedright\begin{thebibliography}{10}

\bibitem{Gaiotto:2009we}
D.~Gaiotto, {\slshape {${\mathcal{N}}\!=2$ Dualities},}
  \href{http://dx.doi.org/10.1007/JHEP08(2012)034}{{\em JHEP} {\bfseries 1208}
  (2012) 034},
\href{http://arxiv.org/abs/0904.2715}{{ arXiv:0904.2715~[hep-th]}}.

\bibitem{Gaiotto:2009hg}
D.~Gaiotto, G.~W. Moore, and A.~Neitzke, {\slshape {Wall-Crossing, Hitchin
  Systems, and the WKB Approximation},}
\href{http://arxiv.org/abs/0907.3987}{{ arXiv:0907.3987~[hep-th]}}.

\bibitem{Chacaltana:2010ks}
O.~Chacaltana and J.~Distler, {\slshape {Tinkertoys for Gaiotto Duality},}
  \href{http://dx.doi.org/10.1007/JHEP11(2010)099}{{\em JHEP} {\bfseries 1011}
  (2010) 099},
\href{http://arxiv.org/abs/1008.5203}{{ arXiv:1008.5203~[hep-th]}}.

\bibitem{Chacaltana:2011ze}
O.~Chacaltana and J.~Distler, {\slshape {Tinkertoys for the $D_{N}$ Series},}
  \href{http://dx.doi.org/10.1007/JHEP02(2013)110}{{\em JHEP} {\bfseries 02}
  (2013) 110},
\href{http://arxiv.org/abs/1106.5410}{{ arXiv:1106.5410~[hep-th]}}.

\bibitem{Chacaltana:2012zy}
O.~Chacaltana, J.~Distler, and Y.~Tachikawa, {\slshape {Nilpotent Orbits and
  Codimension-Two Defects of 6D N=(2,0) Theories},}
  \href{http://dx.doi.org/10.1142/S0217751X1340006X}{{\em Int.J.Mod.Phys.}
  {\bfseries A28} (2013) 1340006},
\href{http://arxiv.org/abs/1203.2930}{{ arXiv:1203.2930~[hep-th]}}.

\bibitem{Chacaltana:2012ch}
O.~Chacaltana, J.~Distler, and Y.~Tachikawa, {\slshape {Gaiotto Duality for the
  Twisted A$_{2N-1}$ Series},}
  \href{http://dx.doi.org/10.1007/JHEP05(2015)075}{{\em JHEP} {\bfseries 05}
  (2015) 075},
\href{http://arxiv.org/abs/1212.3952}{{ arXiv:1212.3952~[hep-th]}}.

\bibitem{Chacaltana:2013oka}
O.~Chacaltana, J.~Distler, and A.~Trimm, {\slshape {Tinkertoys for the Twisted
  D-Series},}
\href{http://arxiv.org/abs/1309.2299}{{ arXiv:1309.2299~[hep-th]}}.

\bibitem{Chacaltana:2014jba}
O.~Chacaltana, J.~Distler, and A.~Trimm, {\slshape {Tinkertoys for the E$_{6}$
  Theory},} \href{http://dx.doi.org/10.1007/JHEP09(2015)007}{{\em JHEP}
  {\bfseries 09} (2015) 007},
\href{http://arxiv.org/abs/1403.4604}{{ arXiv:1403.4604~[hep-th]}}.

\bibitem{Chacaltana:2014ica}
O.~Chacaltana, J.~Distler, and A.~Trimm, {\slshape {Seiberg-Witten for
  $Spin(n)$ with Spinors},}
  \href{http://dx.doi.org/10.1007/JHEP08(2015)027}{{\em JHEP} {\bfseries 08}
  (2015) 027},
\href{http://arxiv.org/abs/1404.3736}{{ arXiv:1404.3736~[hep-th]}}.

\bibitem{Chacaltana:2014nya}
O.~Chacaltana, J.~Distler, and A.~Trimm, {\slshape {A Family of $4D$
  $\mathcal{N}=2$ Interacting SCFTs from the Twisted $A_{2N}$ Series},}
\href{http://arxiv.org/abs/1412.8129}{{ arXiv:1412.8129~[hep-th]}}.

\bibitem{Chacaltana:2015bna}
O.~Chacaltana, J.~Distler, and A.~Trimm, {\slshape {Tinkertoys for the Twisted
  $E_{6}$ Theory},} \href{http://dx.doi.org/10.1007/JHEP04(2015)173}{{\em JHEP}
  {\bfseries 04} (2015) 173},
\href{http://arxiv.org/abs/1501.00357}{{ arXiv:1501.00357~[hep-th]}}.

\bibitem{Chacaltana:2016shw}
O.~Chacaltana, J.~Distler, and A.~Trimm, {\slshape {Tinkertoys for the
  Z3-Twisted D4 Theory},}
\href{http://arxiv.org/abs/1601.02077}{{ arXiv:1601.02077~[hep-th]}}.

\bibitem{Chacaltana:2017boe}
O.~Chacaltana, J.~Distler, A.~Trimm, and Y.~Zhu, {\slshape {Tinkertoys for the
  E7 Theory},}
\href{http://arxiv.org/abs/1704.07890}{{ arXiv:1704.07890~[hep-th]}}.

\bibitem{Distler:2017xba}
J.~Distler, B.~Ergun, and F.~Yan, {\slshape {Product SCFTs in Class-S},}
\href{http://arxiv.org/abs/1711.04727}{{ arXiv:1711.04727~[hep-th]}}.

\bibitem{Chcaltana:2018zag}
O.~Chacaltana, J.~Distler, A.~Trimm, and Y.~Zhu, {\slshape {Tinkertoys for the
  $E_{8}$ Theory},}
\href{http://arxiv.org/abs/1802.09626}{{ arXiv:1802.09626~[hep-th]}}.

\bibitem{Hori:1998iv}
K.~Hori, {\slshape {Consistency Condition for Fivebrane in M-theory on $
  R^5$/$Z_2$ Orbifold},}
  \href{http://dx.doi.org/10.1016/S0550-3213(98)00728-7}{{\em Nucl. Phys.}
  {\bfseries B539} (1999) 35--78},
\href{http://arxiv.org/abs/hep-th/9805141}{{ arXiv:hep-th/9805141}}.

\bibitem{Tachikawa:2011ch}
Y.~Tachikawa, {\slshape {On S-Duality of 5D Super Yang-Mills on S$^1$},}
  \href{http://dx.doi.org/10.1007/JHEP11(2011)123}{{\em JHEP} {\bfseries 11}
  (2011) 123},
\href{http://arxiv.org/abs/1110.0531}{{ arXiv:1110.0531~[hep-th]}}.

\bibitem{Shapere:2008zf}
A.~D. Shapere and Y.~Tachikawa, {\slshape {Central charges of N=2
  superconformal field theories in four dimensions},}
  \href{http://dx.doi.org/10.1088/1126-6708/2008/09/109}{{\em JHEP} {\bfseries
  09} (2008) 109},
\href{http://arxiv.org/abs/0804.1957}{{ arXiv:0804.1957~[hep-th]}}.

\bibitem{Witten:1982fp}
E.~Witten, {\slshape {An $SU(2)$ Anomaly},}
\href{http://dx.doi.org/10.1016/0370-2693(82)90728-6}{{\em Phys. Lett.}
  {\bfseries B117} (1982) 324--328}.

\bibitem{Wang:2015mra}
Y.~Wang and D.~Xie, {\slshape {Classification of Argyres-Douglas Theories from
  M5 Branes},} \href{http://dx.doi.org/10.1103/PhysRevD.94.065012}{{\em Phys.
  Rev.} {\bfseries D94} no.~6, (2016) 065012},
\href{http://arxiv.org/abs/1509.00847}{{ arXiv:1509.00847~[hep-th]}}.

\bibitem{Zafrir:2016wkk}
G.~Zafrir, {\slshape {Compactifications of 5D Scfts with a Twist},}
  \href{http://dx.doi.org/10.1007/JHEP01(2017)097}{{\em JHEP} {\bfseries 01}
  (2017) 097},
\href{http://arxiv.org/abs/1605.08337}{{ arXiv:1605.08337~[hep-th]}}.

\bibitem{Beem:2013sza}
C.~Beem, M.~Lemos, P.~Liendo, W.~Peelaers, L.~Rastelli, and B.~C. van Rees,
  {\slshape {Infinite Chiral Symmetry in Four Dimensions},}
  \href{http://dx.doi.org/10.1007/s00220-014-2272-x}{{\em Commun. Math. Phys.}
  {\bfseries 336} no.~3, (2015) 1359--1433},
\href{http://arxiv.org/abs/1312.5344}{{ arXiv:1312.5344~[hep-th]}}.

\bibitem{wangxie}
Y.~Wang and D.~Xie, {\slshape {Codimension-two defects and Argyres-Douglas
  theories from outer-automorphism twist in 6d $(2,0)$ theories},}
\href{http://arxiv.org/abs/1805.08839}{{ arXiv:1805.08839~[hep-th]}}.

\bibitem{Benini:2009aa}
F.~Benini, S.~Benvenuti, and Y.~Tachikawa, {\slshape {Webs of five-branes and
  N=2 superconformal field theories},}
  \href{http://dx.doi.org/10.1088/1126-6708/2009/09/052}{{\em JHEP} {\bfseries
  09} (2009) 052}, \href{http://arxiv.org/abs/hep-th/0906.0359}{{
  arXiv:hep-th/0906.0359}}.

\bibitem{AH}
O.~Aharony and A.~Hanany, {\slshape {Branes, superpotentials and superconformal
  fixed points},} \href{http://dx.doi.org/10.1016/S0550-3213(97)00472-0}{{\em
  Nucl. Phys.} {\bfseries B504} (1997) 239--271},
  \href{http://arxiv.org/abs/hep-th/9704170}{{ arXiv:hep-th/9704170}}.

\bibitem{AHK}
O.~Aharony, A.~Hanany, and B.~Kol, {\slshape {Webs of (p,q) five-branes,
  five-dimensional field theories and grid diagrams},}
  \href{http://dx.doi.org/10.1088/1126-6708/1998/01/002}{{\em JHEP} {\bfseries
  01} (1998) 002}, \href{http://arxiv.org/abs/hep-th/9710116}{{
  arXiv:hep-th/9710116}}.

\bibitem{Witten:1997sc}
E.~Witten, {\slshape {Solutions of Four-Dimensional Field Theories via M
  Theory},} \href{http://dx.doi.org/10.1016/S0550-3213(97)00416-1}{{\em
  Nucl.Phys.} {\bfseries B500} (1997) 3--42},
\href{http://arxiv.org/abs/hep-th/9703166}{{ arXiv:hep-th/9703166}}.

\bibitem{Gimon:1998be}
E.~G. Gimon, {\slshape {On the M Theory Interpretation of Orientifold Planes},}
\href{http://arxiv.org/abs/hep-th/9806226}{{ arXiv:hep-th/9806226}}.

\end{thebibliography}\endgroup

\end{document}